\documentclass{llncs} 
\usepackage[english]{babel}
\usepackage{latexsym}
\usepackage{amssymb}

\newenvironment{tab}{\begin{tabbing}
MMMMM\=aaa\=aaa\=aaa\=aaa\=aaa\=aaa\= \kill}{\end{tabbing}}

\makeatletter
\def\refmystepcounter#1{\stepcounter{#1}\protect\gdef 
\@currentlabel {\csname p@#1\endcsname \csname 
the#1\endcsname}}
\makeatother
\newcounter {tabnr}

\def\boks  {\mbox{$\Box$}}
\def\bol    {\mbox{$\bullet$}}
\def\Nat   {\mbox{$\mathbb{N}$}}
\def\S #1/{\mbox {\textsl{#1}}}
\def\B #1/{\mbox {\textbf{#1}}}
\def\R #1/{\mbox {\textrm{#1}}}
\def\T #1/{\mbox {\texttt{#1}}}
\def\eps   {{\mbox{$\varepsilon$}}}
\def\IB    {\mbox{$\mathbb{B}$}}
\def\Oh{{\cal O}}

\begin{document}
\setcounter{tabnr}{-1}
\begin {center}
{\Large\bf Kekul\'e Cells for Molecular Computation}\\
\mbox{}\\
W.H. Hesselink $^1$, J.C. Hummelen $^2$, H.T. Jonkman $^2$,\\ 
H.G. Reker $^1$, G.R. Renardel de Lavalette $^1$, M.H. van der Veen $^2$\\
\verb!{w.h.hesselink,j.c.hummelen,harry.jonkman}@rug.nl!,\\
\verb!rekerh@cs.man.ac.uk!, \ 
\verb!g.r.renardel.de.lavalette@rug.nl!,
\verb!marleen.vanderveen@yale.edu!\\
\today\\
$^1$ Computer Science, University of Groningen, The Netherlands\\
$^2$ Molecular Electronics, Zernike Institute of Advanced Materials,\\
University of Groningen, The Netherlands
\end {center}

\begin{abstract}
\noindent
The configurations of single and double bonds in polycyclic
hydrocarbons are abstracted as Kekul\'e states of graphs. Sending a
so-called soliton over an open channel between ports (external nodes)
of the graph changes the Kekul\'e state and therewith the set of open
channels in the graph. This switching behaviour is proposed as a basis
for molecular computation. The proposal is highly speculative but may
have tremendous impact.

Kekul\'e states with the same boundary behaviour (port assignment)
can be regarded as equivalent. This gives rise to the abstraction of
Kekul\'e cells. The basic theory of Kekul\'e states and Kekul\'e cells
is developed here, up to the classification of Kekul\'e cells with
$\leq 4$ ports.
To put the theory in context, we generalize Kekul\'e states to
semi-Kekul\'e states, which form the solutions of a linear system of
equations over the field of the bits 0 and 1. We briefly study
so-called omniconjugated graphs, in which every port assignment of the
right signature has a Kekul\'e state. Omniconjugated graphs may be useful
as connectors between computational elements. We finally investigate
some examples with potentially useful switching behaviour.
\end{abstract}

\section{Introduction}

In the quest for smaller and smaller computational elements, we may
hope to arrive one day at the level of molecules. Controllable
electrical conductance within molecules is the realm of chemistry. 
One proposal in this direction is to use so-called $\pi$-conjugation 
in polycyclic hydrocarbons, as studied in Marleen van der Veen's PhD
thesis ``$\pi$-Logic'' \cite{Vee06}. 

The development of this field of $\pi$-conjugation and $\pi$-logic
needs several abstractions that belong to branches of mathematics like
graph theory and linear algebra. In mathematics, the symbol $\pi$ is
primarily associated with the circumference and the area of the
circle. The term ``conjugation'' has also several connotations in
mathematics. In this paper, conjugation means the constructive interaction
between a pair of neighbouring (carbon-carbon) $\pi$-orbitals, leading to
delocalisation of the electrons.

The basic physical idea is that the configuration of single and double
bonds in certain polycyclic `aromatic' [note: these are
polyunsaturated hydrocarbons, usually referred to as polycyclic
aromatic hydrocarbons (PAHs), but they need not be aromatic in the
strict chemical sense] hydrocarbons influences the electrical
conductivity between points of the molecule, and can be influenced by
electrical signals over channels in the molecule. In other words, the
molecule can serve as a switch.  Since it was Kekul\'e who proposed,
in 1865, the alternating single and double bonds in the benzene ring
(one of the simplest cyclic hydrocarbons), we prefer to associate the
basic ideas to be exposed here with the name Kekul\'e.

The polycyclic hydrocarbons we are considering have boundary atoms
that can serve as \emph{ports} to probe and modify the electronic
properties of the molecules.  The electrical resistance between two
ports is low when there is a path of alternating single and double
bonds between them \cite{YaR02}. By sending a so-called \emph{soliton}
over the alternating path, the single and double bonds along the path
are toggled \cite{HKSS88}. Such toggling of an alternating path may
open or close other alternating paths in the molecule. This is the
switching behaviour alluded to. The toggling can also be done by
chemical means (for example a redox reaction \cite{Ca82,vDMvdVH06}).

In our abstraction of the molecule, the graph of the atoms and bonds
is kept fixed, while it is allowed to change the multiplicities (single or double) of the
bonds. A configuration of bonds such that every
internal node has precisely one double bond is called a \emph{Kekul\'e
state}. Nodes with precisely one
edge to the remainder of the graph are called \emph{ports}. The
\emph{port assignment} of a Kekul\'e state describes the
multiplicities of the bonds at the ports. A pair of ports is called a
\emph{channel}. A channel is called \emph{open} (low resistance) in
a Kekul\'e state if there is an alternating path between its ports.

Another reason for naming the configurations Kekul\'e states is that
they are to represent closed shell molecules (`Kekul\'e structures'),
i.e., molecules in which all electrons are paired, as opposed to
non-Kekul\'e states.

The term `Kekul\'e state' has its chemical equivalent in `resonance
structure' (or `resonance contributor'), with the restriction of being
a system that is closed-shell and without charges. The number of different Kekul\'e states of the graph is a
measure of the stability of the molecule. In this
sense, Kekul\'e states represent all structures that are regarded in
the Valence Bond Theory \cite{Lew16,HeL27,Cou61,ShH03} of molecules
(as opposed to Molecular Orbital Theory, that also includes charged
and open shell configurations as parts for the total quantum
mechanical description of the electronic structure of a molecule).

The Kekul\'e state is a debatable abstraction. The actual
quantum-mechanical state is a weighted superposition of many states,
in which the Kekul\'e states have high weights; e.g., the two Kekul\'e
states of the benzene hexagon are just two components of a single
quantum-mechanical state. This does not matter for the switching
behaviour described, however, because it turns out that all Kekul\'e
states with the same {port assignment} have the same open (closed) channels
between ports, see Theorem \ref{portchannels} in section
\ref{portass}. It follows that the effect of sending a soliton over an
open channel only effects the port assignment, and consists of
toggling the port assignment only at the ports of the channel.

We introduce Kekul\'e cells to capture this behaviour. More precisely,
we introduce a mathematical concept \emph{cell} that captures the
behaviour, and Kekul\'e cells are those that can be obtained from
graphs with single and double bonds in them. 

A serious physical objection is that the Kekul\'e states form a
qualitative characteristic of the state, whereas the precise energy
levels of the various eigen-states are quantitative. This objection
must be dealt with when the qualitative investigations are leading to
actual technical proposals.

Van der Veen et al.\ \cite{VRJH04} have first proposed certain
$\pi$-conjugated systems that can act as `soldering points' for
molecular wires in the sense that linear $\pi$-conjugated pathways
between all ports (`terminals') exist; i.e., `omniconjugated' systems.
Subsequently, $\pi$-logic \cite{Vee06} was proposed as a way to
perform any Boolean operation within certain $\pi$-conjugated
hydrocarbon frameworks, bearing a number of ports. We now formalize
this approach in a rigorous mathematical and topological way.

The application of graph theory to chemistry is not new. One of the
research areas is the search for quantitative structure-property
relationships based on molecular connectivity invariants of the
hydrogen-suppressed chemical graphs, see the overview article
\cite{Pog00}.  The invariants are numerical functions of graphs that
can be related to properties of classes of compounds like, e.g.,
solubility, crystal density, melting point, etc.  In \cite{SSZ95}, it
is proposed that one of these invariants (a generalized Randic index)
for a certain class of hydrocarbons can be interpreted as an energy
functional that depends on the $\pi$-electron density. The paper
\cite{FoR02} goes beyond numerical invariants and studies energy
levels of $\pi$-conjugated systems by means of the symmetry groups of
the graphs. In our application of graph theory, numerical invariants
and the symmetry of the graphs play no roles yet.

\subsection{Overview}

The formalization steps we need are first sketched in section
\ref{formalize}.  In section \ref{matconv}, we present the
mathematical notations that we need. The graph theory needed is
presented in section \ref{graphs}.

In section \ref{sec:kekule}, we present the theory of Kekul\'e states
and Kekul\'e cells, up to the classification of Kekul\'e cells with
four ports. Section \ref{semikekule} contains the theory of
semi-Kekul\'e states. Here it is proved that the number of Kekul\'e
states for a given port assignment is $\leq 2^{e+1-v}$ where $e$ is the
number of edges and $v$ is the number of nodes of the graph.  This
section serves as an aside to provide context for the Kekul\'e states,
but is not really needed for the application.  In section \ref{typeA},
we discuss graphs for which the Kekul\'e cell is as large as possible;
such graphs are said to be omniconjugated. In section \ref {switch}, we
investigate the functionality of cells for switching
behaviour. Conclusions are drawn in section \ref{conclusions}.

\subsection{The formalization steps} \label{formalize}

We now briefly sketch the central concepts of the theory. Precise
definitions are postponed. 

It is natural and conventional to represent molecules by graphs with
the nodes for atoms and the edges for bonds. Polycyclic
polyunsaturated hydrocarbons are usually almost flat, so that their
graphs are planar. We use a subgraph to indicate the set of double bonds 
of a given molecule.

Nodes of the graph that are linked via only one edge to the remainder
of the graph are called \emph{ports}. A \emph{port assignment} is a
set of ports used to indicate the ports that are to be attached by a
double bond. The \emph{Kekul\'e cell} associated to a graph with port
set $P$ is the set of port assignments in $P$ that allow Kekul\'e
states.

A \emph{cell} in $P$ is defined to be an arbitrary set of port
assignments in $P$. A cell is a \emph{Kekul\'e cell} if there is some
graph for which the cell is the set of Kekul\'e port assignments. A
\emph{functional cell} is a cell together with an initial port
assignment and a system of channels to probe and modify the current
port assignment.

\subsection{Mathematical concepts and notations} \label{matconv}

In this paper, a number of mathematical theorems (and lemmas and
corollaries) are proved. The end of such a proof is indicated by the
symbol \boks.  The Theorems 1, 2, and 3 and Corollary 1 have been
verified with the mechanical theorem prover PVS \cite{OSR01} as a
student project (the report and the proof scripts are available on
request).  The symbol \boks\ is also used at the end of the examples
and remarks.

If $S$ is a set, we write $x\in S$ to denote that $x$ is an element of
$S$. We write $\#S$ to denote the number of elements of $S$.

If $S$ and $T$ are sets, we write $S\cup T$ for the \emph{union} of
$S$ and $T$, which is the set of the elements $x$ with $x\in S$ or
$x\in T$ (or both).  The \emph{intersection} $S\cap T$ consists of the
elements $x$ with $x\in S$ and $x\in T$.  We write $S\setminus T$ for
the \emph{difference} of $S$ and $T$, i.e., the set of elements of $S$
that are not in $T$.

A set $S$ is called a \emph{subset} of a set $U$ (notation 
$S\subseteq U$) if every element of $S$ is an element of $U$. 
The \emph{power set} $\S Pow/(U)$ of $U$ is defined as the set of subsets
of $U$.

We write $S\oplus T$ for the \emph{symmetric difference}, which is
$S\oplus T=(S\setminus T)\cup(T\setminus S)$.  It is easy to verify
that $\oplus$ is commutative: $S\oplus T = T\oplus S$ and associative:
$ (S\oplus T) \oplus R = S\oplus (T \oplus R) $, and that the empty
set $\emptyset$ satisfies $\emptyset\oplus S=S$ and $S\oplus
S=\emptyset$. Therefore, $\S Pow/(U)$ with operation $\oplus$ is a
commutative group with neutral element $\emptyset$.

\subsection{Undirected graphs} \label{graphs}

There is no standard terminology in graph theory, cf.\ \cite{GLS88}.
In this note, all graphs are finite, undirected graphs without
isolated nodes, multiple edges, or self-loops.

We formalize our graphs in the following way. The basic constituents
of graphs are nodes (vertices).  An \emph{edge} is defined to be a set
that consists of two distinct nodes. A \emph{graph} is defined to be a
finite set of edges. The \emph{nodes} of a graph are the elements of
its edges.  If $G$ is a graph, we write $nG$ for its set of nodes. The
\emph{ports} of graph $G$ are the nodes that occur in precisely one
edge of $G$ (ports are called sites or terminals in \cite{Vee06}).
We write $pG$ to denote the set of ports of graph $G$.
A node of $G$ is called \emph{internal} if it is not a port. We write
$iG$ to denote the set of internal nodes of $G$. So we have 
$iG=nG\setminus pG$. 

The \emph{degree} of a node $v$ of graph $G$ is the number of edges of
$G$ that contain $v$. Therefore, a node of $G$ is a port if and only
if its degree is 1.  In view of the application to conjugation in
carbon chemistry, we could restrict attention to graphs where all
nodes have degrees $\leq 4$. Since we have no use for this
restriction, we do this only in some examples.

We often represent graphs by drawing the nodes as bullets and the
edges as line segments between the nodes. The graph drawn here has two
ports and three internal nodes, one of degree 2 and two of degree 3.

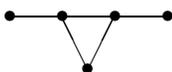
\setlength{\unitlength}{0.7mm}
\begin{picture}(120,18)(-20,2)
\put(15,15.2) {\line(1,0){30}}
\multiput(14.0,14.0)(10,0){4} {\bol}
\put(28.8,4.0) {\bol}
\put(30,5.2) {\line(-1,2){5.3}}
\put(30,5) {\line(1,2){5.3}}
\end{picture}

Two distinct nodes $p$ and $q$ are defined to be \emph{connected} in
graph $G$ if there is a sequence of nodes $(p_0, \dots, p_n)$ with
$p_0=p$ and $p_n=q$ and $(p_i, p_{i+1})\in G$ for all $0\leq i < n$.
Graph $G$ is called \emph{connected} if it is nonempty and every pair
of distinct nodes of $G$ is connected in $G$.  Graph $G$ is called a
\emph{simple path} between nodes $p$ and $q$ if it is connected, $p$
and $q$ are its only ports, and all its other nodes have degree 2.
Graph $G$ is called a \emph{cycle} if it is connected and all its
nodes have degree 2.

Since a graph is just a set of edges, every subset of a graph $G$ is
itself a graph, and can therefore be called a \emph{subgraph}.  A
subgraph $C$ of $G$ is called a \emph{curve} in $G$ if every node of
$C$ which is internal in $G$ has degree 2 in $C$.

The empty subgraph is a curve. More interesting examples of curves are
cycles, and simple paths between ports.  Conversely, every
connected component of a curve is a cycle or a simple path between
ports.

\section {Kekul\'e States} \label{sec:kekule}

The concept of Kekul\'e states is motivated by their correspondence to
$\pi$-conjugated systems in chemistry. Polycyclic polyunsaturated
hydrocarbons are molecules in which carbon atoms are organized in
rings, mainly hexagons, but possibly also pentagons and heptagons.
The primary example would be naphthalene, consisting of two fused
benzene rings.

In 1865, August Kekul\'e proposed that a benzene molecule should
consist of a hexagon of carbon atoms with attached hydrogen in an
alternating cycle of double bonds and single bonds. In our graph
theoretical abstraction, we neglect all hydrogen atoms. All carbon
atoms are represented by nodes of the graph, all bonds are represented
by edges. So, they are the hydrogen suppressed chemical graphs of
\cite{Pog00}.

We represent the set of double bonds by a subgraph of the graph of the
molecule.  In common polycyclic polyunsaturated hydrocarbons, the rule
is that every carbon atom is sp$^2$ hybridized and has precisely one
double bond. The ports represent other atoms that are linked to the
graph by single or double bonds. This leads to the following
definition.

We define a \emph{Kekul\'e state} $W$ in graph $G$ to be a subgraph
of $G$ such that every internal node of $G$ is a port of $W$.

In graph theory, a subgraph $W$ of $G$ is called a \emph{perfect
  matching} if every node of $G$ is a port of $W$ \cite[p.\
203]{GLS88}.  It follows that $W$ is a perfect matching if and only if
$W$ is a Kekul\'e state and $W$ contains all edges to ports.

\begin{example}
  The graph shown in section \ref{graphs} has no perfect matchings,
  and has two Kekul\'e states, one of which ($\{e,f\}$) is shown below
  by doubling its edges $e$ and $f$. If we give the bottom node an
  additional edge to a new port, the resulting graph has one perfect
  matching and four Kekul\'e states.

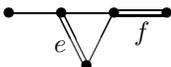
\begin{picture}(120,18)(-20,2)
\put(15,15.2) {\line(1,0){20}}
\multiput(35,14.7)(0,1){2} {\line(1,0){10}}
\multiput(14.0,14.0)(10,0){4} {\bol}
\put(28.8,4.0) {\bol}
\multiput(29.5,5.2)(1,0){2} {\line(-1,2){5.1}}
\put(23.8,7.8) {$e$}
\put(39,10) {$f$}
\put(30,5) {\line(1,2){5.3}}
\end{picture}

\end{example}

Efficient algorithms exist to determine whether a given graph
has a perfect matching \cite{MiV80}. These algorithms can easily be
adapted to yield Kekul\'e states. In the remainder of this paper, 
we have no need to consider perfect matchings.

\begin{remark} There are graphs, even with several ports, that have no
Kekul\'e states. The lefthand graph below is an example, as is easily
verified. In this graph, the lower part with the two ports can be
replaced by any other subgraph with any number of ports.

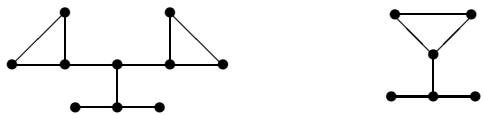
\begin{picture}(120,30)(6,0)
\put(20,15) {\line(1,0){40}}
\multiput(18.8,13.8)(10,0){5} {\bol}
\multiput(28.8,23.8)(20,0){2} {\bol}
\multiput(30.8,5.8)(8,0){3} {\bol}
\multiput(30,15)(20,0){2} {\line(0,1){10}}
\put(20,15) {\line(1,1){10}}
\put(60,15) {\line(-1,1){10}}
\multiput(40,15)(60,2){2} {\line(0,-1){8}}
\multiput(32,7)(60,2){2} {\line(1,0){16}}
\put(98.8,15.8) {\bol}
\multiput(90.8,7.8)(8,0){3} {\bol}
\multiput(91.5,23.5)(14.5,0){2} {\bol}
\put(100,17) {\line(1,1){7}}
\put(100,17) {\line(-1,1){7}}
\put(93,24.7) {\line(1,0){14}}
\end{picture}

\noindent 
The righthand graph in the figure is a case with precisely one
Kekul\'e state, even though the graph has two ports. In either case,
the triangles can be replaced by pentagons to better meet the
possibilities of carbon chemistry. 

The chemical equivalents of the two structures are depicted below. The
structure at the left is an open shell (non-Kekul\'e) radical
structure with the bullet at a Kekul\'e violation. The
structure at the right is a closed shell (Kekul\'e) structure.

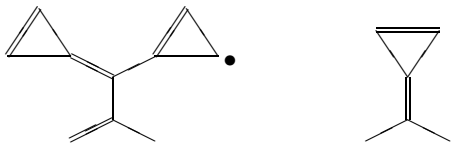
\begin{picture}(120,30)(0,18)
\multiput(0,0)(28,0){2}{
 \put(20,35){\line(1,0){12}}
 \multiput(19.7,35.5)(0.6,-0.6){2}{\line(2,3){6}}
 \put(32,35){\line(-2,3){6}}
}
\multiput(32,34.7)(0.2,0.7){2}{\line(2,-1){8}}
\put(48,35){\line(-2,-1){8}}
\multiput(32,18.7)(-0.2,0.7){2}{\line(2,1){8}}
\put(48,19){\line(-2,1){8}}
\put(40,31){\line(0,-1){8}}
\put(61,33){\bol}
\put(90,40){\line(2,-3){6}}
\put(102,40){\line(-2,-3){6}}
\multiput(90,39.6)(0,0.8){2}{\line(1,0){12}}
\multiput(95.6,31)(0.8,0){2}{\line(0,-1){8}}
\put(88,19){\line(2,1){8}}
\put(104,19){\line(-2,1){8}}
\end{picture}

\end{remark}

\subsection{Alternating curves}

The electronic properties of Kekul\'e states are observed at and
controlled by alternating paths, or more generally alternating curves. 

An \emph{alternating curve} is a pair $(C,W)$ of subgraphs of $G$ such
that $C$ is a curve in $G$ and that the intersection $W\cap C$ is a
Kekul\'e state of $C$.  Since all internal nodes of $C$ have degree 2,
this means that every internal node of $C$ belongs to one edge
of $W$ and one edge of $C\setminus W$, whence the name
``alternating''.

Roughly speaking, the next result says that the difference between any
two Kekul\'e states is an alternating curve, and that, if at
least one Kekul\'e state exists, every alternating curve is
obtained in this way.

\begin{theorem} \label{difww} (a) Let $W$ and $W'$ be Kekul\'e states
  of a graph $G$. Then the sub\-graph $C=W\oplus W'$ is a curve in $G$
  and $(C,W)$ is an alternating curve.\\ 
  (b) Let $W$ be a Kekul\'e state of graph $G$ and let $(C,W)$ be an
  alternating curve in $G$. Then $W'=W\oplus C$ is a Kekul\'e state of
  $G$.
\end{theorem}

\begin{proof}
  
  (a) Let $v$ be a node of $C$ which is an internal node of $G$.
  Since $W$ and $W'$ are Kekul\'e states of $G$, each contains
  precisely one edge, say $e\in W$ and $e'\in W'$, that contains
  $v$. Since $v$ is a node of $C=W\oplus W'$, these edges are
  different. Therefore, both belong to $C$. It also follows that $C$
  has no other edges that contain $v$.  Therefore $v$ has degree 2 in
  $C$. This proves that $C$ is a curve in $G$. Moreover, since we have
  $e\in W\setminus W'$ and $e'\in W'\setminus W$, the pair $(C,W)$ is
  alternating.

  (b) In order to show that $W'$ is a Kekul\'e state, we need to show
  that every internal node $v$ of $G$ is a port of $W'$. Since $W$ is
  a Kekul\'e state of $G$, there is a unique edge $e\in W$ with $v\in
  e$. If $v$ is not a node of $C$, then $e$ is also the unique edge of
  $W\oplus C$ that contains $v$.  Assume therefore that $v$ is a node
  of $C$. Since $C$ is a curve, $v$ is internal in $C$ and has degree
  2 in $C$.  Since $(C,W)$ is an alternating curve, it follows that
  there is a unique edge $e\in W\cap C$ and a unique edge $e'\in
  C\setminus W$ with $v\in e$ and $v\in e'$. It follows that $e'$ is
  the unique edge of $ W\oplus C=W'$ that contains $v$. Therefore $v$
  is a port of $W'$. \boks
\end{proof}

\subsection {Port assignments} \label{portass}

We now want to prescribe the port behaviour of Kekul\'e states.  Let
$G$ be a graph with a set of ports $P$.  We define a \emph{port
  assignment} (``archetype'' in \cite{Vee06}) to be a subset of $P$.
For any subgraph $W$ of $G$, we define $W|P$ as the set of nodes of
$W$ that also belong to $P$, i.e.,\ $(W|P)= P\cap nW$. A port
assignment $g$ is called \emph{Kekul\'e} if there is a Kekul\'e state
$W$ of $G$ with $g=(W|P)$.

In chemistry, the different Kekul\'e states with the same port
assignment are called resonance structures. The actual quantum
mechanical state is a superposition of these resonance structures.  In
the Valence Bond Theory picture, the actual quantum mechanical state
is taken as a weighted superposition of these (neutral) resonance
structures.  The number of neutral closed shell resonance structures
is determined by the number of alternating curves in the following
way:

\begin{corollary}
  Let $G$ be a graph with $pG=P$. Let $g$ be a port assignment in $P$.
  Let $n$ be the number of Kekul\'e states $W$ of $G$ with $(W|P)=g$.
  For every Kekul\'e state $W$ with $(W|P)=g$, there are precisely $n$
  curves $C$ without ports such that $(C,W)$ is alternating (including
  the empty curve).
\end{corollary}

\begin{proof}
This follows from Theorem \ref{difww} by the observation that $W$ and $W'$
give the same port assignments if and only if the curve $W\oplus W'$
has no ports. \boks
\end{proof}

\begin{example} Phenantrene consists of three hexagons without any ports. 
One of its Kekul\'e states is depicted here. 

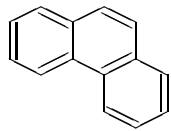
\begin{picture}(120,30)(0,17)
\multiput(20,40)(12,0){2} {\line(2,1){6}}
\multiput(26,43)(12,0){2} {\line(2,-1){6}}
\multiput(25.5,41.7)(12,0){2} {\line(2,-1){5.7}}
\multiput(20,40)(12,0){3} {\line(0,-1){7}}
\multiput(20,33)(12,0){3} {\line(2,-1){6}}
\multiput(26,30)(12,0){2} {\line(2,+1){6}}
\multiput(25.5,31.3)(12,0){2} {\line(2,+1){5.7}}
\put(21.3,40)  {\line(0,-1){7}}
\multiput(38,30)(12,0){2} {\line(0, -1){7}}
\multiput(38,23)(.5,1){2} {\line(2,-1){6}}
\put(50,23)    {\line(-2,-1){6}}
\put(48.7,23)    {\line(0,1){7}}

\end{picture}

\noindent
This Kekul\'e state has five alternating curves. Four of them are easy
to find: the empty curve, two cycles of length 6 around the extreme
hexagons, and one cycle of length 10 around the two horizontal
neighbouring hexagons.  The remaining alternating curve is
disconnected: it is the union of the two cycles of length 6.
Phenantrene has therefore five Kekul\'e states. \boks
\end{example}

\medbreak In chemistry, i.e., within Valence Bond Theory, the number
of Kekul\'e states of a port assignment is an indication of its
stability. Here the Kekul\'e structures having the highest number of
aromatic rings (rings with $4n + 2$ electrons; usually 6, as in
benzene) are considered as the most important resonance contributors.

We now come to the central result of this paper that says that the
existence of an alternating path between two given ports is completely
determined by the set of Kekul\'e port assignments. It follows that
the switching behavour of the graph is also completely determined by
the set of Kekul\'e port assignments and hence independent of all
other aspects of the graph.

In order to formulate this result, we note that, for distinct ports
$p$ and $q$, the doubleton set $\{p,q\}$ is a port assignment.

\begin{theorem} \label{portchannels}
Let port assignment $k$ satisfy $k=(W|P)$ for some Kekul\'e state
$W$. Let $p$ and $q$ be distinct ports. There is a simple alternating
path $(C,W)$ from $p$ to $q$ in $G$ if and only if the port assignment
$k\oplus\{p,q\}$ is Kekul\'e.
\end{theorem}

\begin{proof}
If there is a simple alternating path $(C,W)$ from $p$ to $q$ in $G$,
Theorem \ref{difww}(b) implies that $W'=W\oplus C$ is a Kekul\'e
state of $G$. It satisfies $(W'|P) = k\oplus \{p,q\} $.

Conversely, if the port assignment $k\oplus \{p,q\} $ is Kekul\'e, let
$W'$ be a Kekul\'e state with $(W'|P) = k\oplus \{p,q\} $ and let
$C'=W\oplus W'$. Then $(C',W)$ is an alternating curve by Theorem
\ref{difww}(a). The Kekul\'e states $W$ and $W'$ agree on all ports
except for $p$ and $q$. Therefore, the graph $C'$ only contains the
ports $p$ and $q$. Let $C$ be the connected component of $p$ in
$C'$. Then it also contains $q$ and $(C,W)$ is a simple alternating
path from $p$ to $q$ in $G$. \boks
\end{proof}

\medbreak 
Theorem \ref{portchannels} implies that all Kekul\'e states
corresponding to a given Kekul\'e port assignment agree on the
question which ports are connected by alternating paths, and that the
result is determined completely by the set of Kekul\'e port
assignments. 

The physical relevance of Theorem \ref{portchannels} is that an
alternating path between a pair of ports is, electrically, an open
channel (low resistance), and that the channel is closed in the absence of alternating paths \cite{NiR03,MWREao03,DMKao03}. This leads to the
following formal definitions.

In view of Theorem \ref{portchannels}, we define a \emph{channel} to
be a port assignment of the form $\{p,q\}$ for distinct ports
$p$ and $q$. A channel $c$ is defined to be \emph{open} with respect to a
set $K$ of port assignments and an element $k\in K$ if and only if
$k\oplus c\in K$.

The idea is that $K$ stands for the set of Kekul\'e port assignments
of some graph. If a channel $c$ is open for $k\in K$, one can send
a so-called soliton through the channel with the effect that state $k$
is replaced by $k\oplus c$ \cite{Vee06}.  Indeed, the element $k\in K$
is regarded as a memory state. It is up to the physics to determine
how long such memory states can be preserved.
\begin{example}
  One of the simplest cases is an ethene molecule with three attached
  ports $p_0$, $p_1$, $p_2$. The initial port assignment is $\emptyset$.
  We use the channels $A = \{p_0,p_1\}$ and $T = \{p_0, p_2\}$.  There
  are three Kekul\'e states, as depicted below. The corresponding port
  assignments are $\emptyset$, $A$ and $A\oplus T = \{p_1,p_2\}$. Port
  assignment $T$ is non-Kekul\'e.

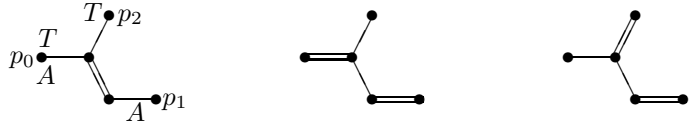
\setlength{\unitlength}{0.7mm}
\begin{picture}(150,30)(15,13)
\multiput (0,0)(50,0){3}
  {\multiput(0,0)(12.8, -8){2}{\multiput(32.8,26.8)(9,0){2}{$\bullet$}}
   \put(45.6,34.8){$\bullet$}}
\put(46.5,20) {\line(1,0){9}}
\multiput (0,0)(100,0){2} {\put(34,28) {\line(1,0){9}}}
\multiput (0,0)(50,0){2} {\put(43,28) {\line(1,2){4}}}
\multiput(42.5,28)(1,0){2} {\line(1,-2){4}}
\multiput(93,28)(50,0){2} {\line(1,-2){4}}
\multiput(0,0)(50,0){2}{\multiput(97.5,19.5)(0,1){2} {\line(1,0){9}}}
\multiput(84.5,27.6)(0,1){2} {\line(1,0){8}}
\multiput(142.4,28)(1,0){2} {\line(1,2){3.8}}
\put(28, 27) {$p_0$} 
\put(33, 23) {$A$}
\put(33.5, 30) {$T$}
\put(57, 19) {$p_1$}
\put(48.5, 35) {$p_2$}
\put(42, 35) {$T$}
\put(50, 16) {$A$}
\end{picture}

\noindent
Initially, channel $A$ is open and $T$ is closed. When $A$ is
signalled, we arrive at the middle state: this opens $T$, while $A$
remains open. If $A$ is signalled again, we return to the initial
state where $T$ is closed. We may use the system as a simple switch to
open and close $T$ by means of $A$. Note, however, that if $A$ opens
$T$, and $T$ is then read, i.e., is signalled, we arrive at the
righthand state where channel $A$ is closed.  Therefore, in order to
preserve the system as a switch, reading this memory cell must be done
by signalling the channel twice.  \boks
\end{example}

\subsection{Abstraction from the graphs} \label{abstr}

At this point we take the abstraction step to concentrate on the ports
and the port assignments and treat the specific graph itself merely
as auxiliary. From here on, $P$ is an arbitrary finite set and the elements of
$P$ are called \emph{ports}. The subsets of $P$ are called
\emph{port assignments}.  Just as before, a \emph{channel} is a port
assignment of the form $\{p,q\}$ for distinct ports $p$ and $q$.

Recall from section \ref{matconv} that the set $\S Pow/(P)$ of the port 
assignments is a commutative group with operation $\oplus$. 
We define $\S Even/(P)$ and $\S Odd/(P)$ to be the subsets of $\S Pow/(P)$
that consist of the port asignments with an even (odd) number of ports.
It is easy to verify that $\S Even/(P)$ is a subgroup of $\S Pow/(P)$
and that $\S Odd/(P)$ is not.

For $\#P=4$, the sets $\S Even/(P)$ and $\S Odd/(P)$ are depicted in
Figure 5.5 of \cite{Vee06}.  Table 5.10 of \cite{Vee06} gives the
table of the operation $\oplus$ in $\S Even/(P)$.  The group $\S
Even/(P)$ \emph{acts} on $\S Odd/(P)$ in the sense that $g\oplus h\in
\S Odd/(P)$ for all $g\in\S Even/(P)$ and $h\in\S Odd/(P)$. All
channels are elements of $\S Even/(P)$.

For any two port assignments $k$, $k'\in \S Pow/(P)$, we define the
\emph{Hamming distance} $\S dist/(k,k')$ as the number of ports in $
k\oplus k'$. It is easy to verify that $\S dist/(k,k')$ is always $\geq
0$, and that it is zero if and only if $k=k'$. Recall that, in any
group, \emph{translation} is the operation that adds a fixed element
to its argument. The Hamming distance is translation invariant: $\S
dist/(g\oplus k, g\oplus k')=\S dist/(k,k')$ for all $g$, $k$,
$k'\in\S Pow/(P)$.

\subsection{Cells and Kekul\'e cells} \label{Kekport}

A \emph{cell} over $P$ is defined to be a subset of $\S Pow/(P)$. A
channel $c$ is defined to be \emph{open} with respect to a cell $K$
and a state $k\in K$, if and only if $k\oplus c\in K$. In accordance
with Theorem \ref{portchannels}, the interpretation is that a signal
can be sent over channel $c$ if and only if the channel is open and
that, in that case, the state is transformed into $k\oplus c$.

For any graph $G$, we let $\S KP/(G)$ denote the set of the Kekul\'e
port assignments of $G$. Clearly $\S KP/(G)$ is a cell over $P$, i.e.,
a subset of $\S Pow/(P)$.  An arbitrary cell $K$ over $P$ is called a
\emph{Kekul\'e cell} if and only if there is a graph $G$ with $P=pG$
and $K=\S KP/(G)$.

At this point the question is: how arbitrary are Kekul\'e cells?  Our
first result in this direction is that Kekul\'e cells are preserved by
translation.  If $K$ is a cell and $g$ is a port assignment, the
\emph{translated cell} $g\oplus K$ is defined by $g\oplus K=\{g\oplus
k\mid k\in K\}$.

\begin{theorem} \label{kektranslated} Let $K$ be a Kekul\'e cell over
  $P$. Let $g\in \S Pow/(P)$. Then $g\oplus K$ is also a Kekul\'e cell
  over $P$.
\end{theorem}

\begin{proof} 
  Choose a graph $G$ with $P = pG$ and $K=\S KP/(G)$. It suffices to
  construct a graph $G'$ with $g\oplus K=\S KP/(G')$. We construct
  $G'$ from $G$.  For every port $p\in g$, we replace the (unique)
  edge of $G$ that contains $p$ by two edges linked by a new internal
  node.  More precisely, let each edge $e=\{p,v\}$ with $p\in g$ be
  replaced by the edges $\{p,v'\}$ and $\{v',v\}$.  If $W$ is a
  Kekul\'e state of $G$, let $W'$ be the subgraph of $G'$ that
  consists of $W\cap G'$ together with all edges $\{v',v\}$ with $p\in
  g$ and $\{p,v\}\in W$, and all edges $\{p,v'\}$ with $p\in g$ and
  $\{p,v\}\in G\setminus W$. Then $W'$ is a Kekul\'e state of $G'$
  with $W'|P = g\oplus (W|P)$.  Moreover, every Kekul\'e state $W'$ of
  $G'$ is obtained in this way. This shows that $g\oplus K=\S
  KP/(G')$. \boks
\end{proof}

\medbreak We now prove a result that shows that every Kekul\'e cell is
``connected by channels'' in a certain, rather strong, sense.  For any
set $D$ of port assignments we use the notation $\bigoplus D$ to
denote the $\oplus$ sum of the elements of $D$.

\begin{theorem} \label {kekConByChan} Let $K$ be a Kekul\'e cell over
  $P$. For every pair of elements $g$, $g'\in K$, the Hamming distance
  $\S dist/(g,g')$ is even and there is a set $D$ of disjoint channels
  such that $g'=g\oplus (\bigoplus D)$ and $\S dist/(g,g')=2\cdot \#
  D$ and, in addition, $ g\oplus(\bigoplus D')\in K$ for every subset
  $D'$ of $D$.
\end{theorem}

\begin{proof}
Since $K$ is a Kekul\'e cell, we can choose a graph $G$ with $P=pG$
and $K=\S KP/(G)$.  Let $g$, $g'\in K$. We can take Kekul\'e states
$W$ and $W'$ of $G$ so that $g=(W|P)$ and $g'=(W'|P)$. Let $(C,W)$ be the
alternating curve of Theorem \ref{difww}(a) with $C= W\oplus W'$.
Curve $C$ is the disjoint union of simple paths between ports, and
cycles. Assume that $C_0$, \dots, $C_{d-1}$ are the components
of $C$ that are the simple paths.  Then $(C_i,W)$ is an alternating
path between ports for every $i<d$.

  For each index $i$, let $c_i$ be the channel that consists of the
  endpoints of path $C_i$. We take $D=\{c_i\mid i < d\}$. Since the
  paths $C_i$ are components of $C$, they are disjoint. Therefore, the
  channels in $D$ are disjoint. It is easy to see that $g'=g\oplus
  (\bigoplus D)$ and $\S dist/(g,g')=2\cdot \# D$. Repeated application
  of Theorem \ref{difww}(b) yields that $ g\oplus(\bigoplus D')\in \S
  KP/(G) = K $ for every subset $D'$ of $D$. \boks
\end{proof}

\medbreak This result implies that, in a Kekul\'e cell, every two
elements differ by a set of disjoint channels in such a way that all
intermediate combinations (of one element with some of the channels)
also belong to the cell.

A port $p\in P$ is called \emph{flexible for a cell} $K$ if there are
$g$, $g'\in K$ with $p\in g\setminus g'$. Nonflexible ports do not
participate in any switching behaviour. We are therefore only
interested in the flexible ports.  Now, Theorem \ref{kekConByChan}
also implies the following result.

\begin{corollary} \label{flexprop}
Let $K$ be a Kekul\'e cell in $\S Pow/(P)$.\\
(a) Let $p$ be a flexible port for $K$. For any $g\in K$, there is
a channel $c$ with $p\in c$ and $g\oplus c\in K$.\\
(b) When $K$ has a flexible port, it has to have at least two flexible ports. 
\end{corollary}

Since nonflexible ports are useless for the application we have in
mind, we define a cell to be \emph{flexible} if all ports are flexible.
We eliminate the nonflexible ports in the following way. Let $Q$ be the set of
flexible ports of $K$. Then the \emph{flexible cell} $\S flex/(K)\subseteq
\S Pow/(Q)$ is defined by $\S flex/(K)= \{ g\cap Q\mid g\in K\}$.

It is easy to see that, indeed, cell $\S flex/(K)$ is flexible, i.e.,
all ports $p\in Q$ are flexible for $\S flex/(K)$. The restriction
function $g\mapsto g\cap Q$ is a bijective correspondence between $K$
and $\S flex/(K)$.

\begin{theorem}
The cell $\S flex/(K)$ is Kekul\'e if and only if the cell $K$ is Kekul\'e.
\end{theorem}

\begin{proof}
First assume that cell $K$ is Kekul\'e. We can therefore choose a graph $G$
with $P=pG$ and $K=\S KP/(G)$. We define an edge of graph $G$ to be
flexible if there are Kekul\'e states $W$ and $W'$ of $G$ with $e\in
W\setminus W'$. Let $G'$ be the subgraph of the flexible edges of $G$.

We will show that $\S flex/(K)$ is a Kekul\'e cell by demonstrating that $\S
flex/(K)=\S KP/(G')$.  We first need to demonstrate that the ports of $G'$
are the flexible ports of $K$.  If node $v$ of $G$ is a port of $G'$,
it is incident with precisely one edge $e\in G'$; then there are
Kekul\'e states $W$ and $W'$ of $G$ with $e\in W\setminus W'$; if node
$v$ is internal in $G$, it follows that there is also an edge $e'$
that contains $v$ such that $e'\in W'\setminus W$, and hence $e' \in
G'$, so that $v$ is not a port of $G'$, a contradiction.  This shows
that $v$ is a flexible port of $\S KP/(G)=K$. Conversely, every
flexible port of $\S KP/(G)$ clearly is a port of $G'$.

We now claim that, for every Kekul\'e state $W$ of $G$, the
intersection $W\cap G'$ is a Kekul\'e state of $G'$.  Let $v$ be an
internal node of $G'$. Then it is an internal node of $G$ and hence
contained in a unique edge $e\in W$. If $e\notin G'$, then $e\in W'$
for all Kekul\'e states $W'$ of $G$. Since $v$ is contained in a
flexible edge of $G$, this results in a contradiction. This proves that
$e\in G'$ and, therefore, that $W\cap G'$ is a Kekul\'e state
of $G'$. This implies that $\S flex/(K)\subseteq \S KP/(G')$.

Conversely, every Kekul\'e state $W_0$ of $G'$ is the restriction of a
Kekul\'e state $W$ of $G$. In fact, choose an arbitrary Kekul\'e state
$W_1$ of $G$. Define $W$ on $G$ by $W = W_0 \cup (W_1\setminus G')$.
Since, as shown in the first paragraph,
$e\notin W_0$ for every edge $e\notin G'$ that is incident with a node
$v$ of $G'$, the subgraph $W$ is a Kekul\'e state of $G$.  This
implies that $\S flex/(K)\supseteq \S KP/(G')$ and, hence $\S
flex/(K)= \S KP/(G')$.  Therefore, $\S flex/(K)$ is a Kekul\'e cell.

\smallbreak
Conversely, assume that $\S flex/(K)$ is a Kekul\'e cell, say $\S flex/(K)=\S
KP/(G')$ for some graph $G'$. We construct a graph $G$ with $K=\S
KP/(G)$ by extending $G'$ with handles for the nonflexible ports.  The
nonflexible ports fall in two classes. Let $P_1$ consist of the ports
$p\in P$ with $p\in g$ for all $g\in K$ and let $P_0$ be the remainder
$P \setminus (Q\cup P_1)$.

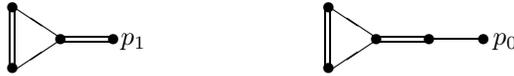
\begin{picture}(120,20)(6,0)
\multiput(0,0)(60,0){2}
{\multiput(20.8,4.2)(1.0,0){2} {\line(0,1){11.8}}
\multiput(29.8,9.5)(0,1.0){2} {\line(1,0){10}}
\multiput(20.2,3.3)(0,11.5){2} {\bol}
\multiput(29.3,8.8)(10,0){2} {\bol}
\put(22,4.2) {\line(3,2){8}}
\put(22,16) {\line(3,-2){8}}}
\put(42,9) {$p_1$}
\put(100,10) {\line(1,0){10}}
\put(112.7,9) {$p_0$}
\put(109.6,8.8) {\bol}
\end{picture}

\noindent
In the diagram, we depict the handles to be attached to the nonflexible 
ports $p_1\in P_1$ and $p_0\in P_0$. The verification that $K=\S KP/(G)$ 
is straightforward. \boks
\end{proof}

\subsection{Transforming the graph while preserving its Kekul\'e cell}

The next result allows us to simplify graphs while retaining the
Kekul\'e cell.

\begin{theorem} \label{mergenodes} Let graph $G$ have a node $u_0$
  with precisely two neighbour nodes $u_1$ and $u_2$. Assume that
  $u_1$ and $u_2$ are both internal. Let graph $G'$ be obtained from $G$ by
  removing $u_0$ and its two incident edges and merging the nodes
  $u_1$ and $u_2$ into a new node $u$.  If there is an edge between
  $u_1$ and $u_2$, it is removed since it would become a self-loop.
  For every node $v\ne u_0$ that is a common neighbour of $u_1$ and
  $u_2$, the two edges linking $v$ to $u_1$ and $u_2$ are identified,
  since multiple edges are not allowed.

  Consequently, the degree of $u$ is $d_1+d_2 - 2 - 2a - b$, where
  $d_1$ and $d_2$ are the degrees of $u_1$ and $u_2$, respectively,
  and $a\in\{0,1\}$ is the number of edges between $u_1$ and $u_2$,
  and $b\in\Nat$ is the number of common neighbours of $u_1$ and $u_2$
  different from $u_0$. In all cases, $\S KP/(G')=\S KP/(G)$.
\end{theorem}

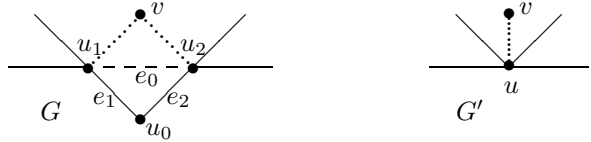
\begin{picture}(120,30)(6,0)
\put(20,25){\line(1,-1){20}}
\put(40,5){\line(1,1){20}}
\multiput(38.8,3.7)(0,20){2}{\bol}
\multiput(28.8,13.5)(20,0){2}{\bol}
\multiput(42,25)(70,0){2}{$v$}
\multiput(39.4,24.5)(-1,-1){11}{.}
\multiput(39.4,24.5)(1,-1){11}{.}
\put(21,5){$G$}
\put(100,5){$G'$}
\put(41,2){$u_0$}
\put(28,18){$u_1$}
\put(47.6,18){$u_2$}
\put(39,12){$e_0$}
\put(31,8){$e_1$}
\put(45,8){$e_2$}
\multiput(33,15)(4,0){4}{\line(1,0){2}}
\multiput(15,15)(35,0){2}{\line(1,0){15}}
\put(100,25){\line(1,-1){10}}
\put(95,15){\line(1,0){30}}
\put(120,25){\line(-1,-1){10}}
\multiput(108.8,14)(0,10){2}{\bol}
\multiput(109.4,15)(0,1){11}{.}
\put(109,10){$u$}
\end{picture}

\begin{proof} 
  Let $e_1$ and $e_2$ be the edges linking $u_0$ with $u_1$ and $u_2$,
  respectively. If it exists, let $e_0$ be the edge between $u_1$ and
  $u_2$. Let $E_1$ be the set of the other edges incident with $u_1$
  and let $E_2$ be the set of the other edges incident with $u_2$.
  Every Kekul\'e state $W$ of $G$ has $e_1\in W$ if and only if
  $e_2\notin W$.  It follows that $W$ contains precisely one of the
  edges in $E_1\cup E_2$ and that $e_0\notin W$ if $e_0$ exists. We can
  therefore transform $W$ into a Kekul\'e state of $G'$ with the same
  values on all remaining edges.  The details about neighbour $v$ are
  left to the reader.  It is easy to see that all Kekul\'e states of
  $G'$ are obtained in this way. Because the transformation does not 
  change the ports, the Kekul\'e cells of $G$ and $G'$ are equal. \boks
\end{proof}

\medbreak This result can also be used in the other direction, i.e.,
starting with graph $G'$.  In this way, it can serve to split a node
with a high degree into two nodes with a lower degree. By first merging
and then splitting, it can also be used to shift edges between
nodes $u_1$ and $u_2$ in a similar way as with operation \B Op.vii/ of 
\cite[section 3.3.1]{Vee06}.

\subsection{Classification of small Kekul\'e cells}

We define the \emph{Hamming diameter} of a cell $K$ as the maximum
Hamming distance between its elements: $\S diam/(K) = \R max/\{\S
dist/(k,k')\mid k,k'\in K\}$. We clearly have $\S diam/(K)\leq
\#P$. Cell $K$ has flexible ports if and only if $\S diam/(K) > 0$. By
Theorem \ref{kekConByChan}, the diameter of a Kekul\'e cell is always
even. We use these observations here to determine all flexible
Kekul\'e cells with at most 4 ports. 

For a fixed set $P$ of ports, we define the cell $ K_1 = \{\{p\}\mid p\in
P\}$.  If $P$ has more than one element, $K_1$ has diameter 2.

\begin{lemma} \label{cellDiam2}
Let $K$ be a flexible cell in $\S Pow/(P)$ with diameter 2. Assume that
$\#P\ne 3$. Then $K$ is a Kekul\'e cell if and only if there is $g\in
\S Pow/(P)$ such that $K=g\oplus K_1$.
\end{lemma}

\begin{proof}
We first show that $K_1$ is a Kekul\'e cell. For this purpose, we take
the graph $G$ with one internal node connected to all ports (a
so-called star). Every Kekul\'e state $W$ of $G$ contains precisely
one edge. The corresponding port assignment contains precisely one
port. Since Kekul\'e cells can be translated, this proves the
\emph{if} part of the assertion.

Conversely, let $K$ be a Kekul\'e cell with diameter 2. By
translation, we may assume that $\emptyset\in K$. By Corollary
\ref{flexprop}(a) with $g = \emptyset$, for every $p\in P$, there is
some $q\ne p$ with $\{p,q\}\in K$. Using $\#P\ne 3$ and $\S diam/(K)=2$,
one can show that there is one port $p_0$ such that $K$ consists of the 
channels $\{p_0,q\}$ for all ports $q\ne p_0$. This implies that
$K=\{p_0\}\oplus K_1$. \boks
\end{proof}

\begin{remark}
A star graph with $> 4$ ports is not commonly found in carbon
chemistry, but by (repeated) application of Theorem \ref{mergenodes}, 
one can easily find alternatives.
\noindent An example where one can use a tree graph for the five port situation is given below. \boks

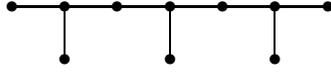
\begin{picture}(120,20)(6,0)
\put(15,15) {\line(1,0){60}}
\multiput(25,5)(20,0){3} {\line(0,1){10}}
\multiput(13.8,13.7)(10,0){7} {\bol}
\multiput(23.8,3.7)(20,0){3} {\bol}
\end{picture} 

\end{remark}

\begin{remark}
In the case of $\#P=3$, all flexible Kekul\'e cells have diameter 2.
In this case, there are two classes of flexible Kekul\'e cells: the
translates of $K_1$ of Lemma \ref{cellDiam2} and the translates of $\S
Even/(P)$. \boks
\end{remark}

\begin{lemma} \label{class4}
Let $P$ have 4 ports. Let $K$ be a cell in $\S Pow/(P)$.  Then $K$ is a
Kekul\'e cell with diameter 4 if and only if there is $g\in\S Pow/(P)$ and
an enumeration $a$, $b$, $c$, $d$ of the ports such that
$$ \{\emptyset, \{a, b\}, \{c, d\}, \{a, b, c, d\} \}\subseteq 
g\oplus K\subseteq\S Even/(P) \: .$$
\end{lemma}

\begin{proof}
First, assume that $K$ is a Kekul\'e cell with diameter 4. By
translation, say over $g$, we may assume that $\emptyset\in K$ and $\S
dist/(\B 0/,k)=4$ for some $k\in K$. By Theorem \ref{kekConByChan},
there are two disjoint channels $\{a, b\}$ and $\{c, d\}$ such that
$K$ contains the port assignments $ \emptyset$, $\{a, b\}$, $\{c,
d\}$, and $\{a, b, c, d\}$.  Also, by Theorem \ref{kekConByChan}, all
elements of $K$ have even distance to $\emptyset$. Therefore, $K$ is
contained in $\S Even/(P)$.

For the converse implication (only if), we may assume that $K$
satisfies $K_0\subseteq K\subseteq \S Even/(P)$ where $K_0=
\{\emptyset, \{a, b\}, \{c, d\}, \{a, b, c, d\} \}$.  The only
elements for which it is not known whether they are contained in $K$,
are $\{a, c\}$, $\{a, d\}$, $\{b, c\}$, $\{b, d\}$.  This leaves
$2^4=16$ possibilities for $K$. Using symmetry, we can reduce it to 6
cases, viz.\ $K_1=K_0\cup\{\{a, c\}\}$, $K_2=K_1\cup\{\{b, d\}\}$,
$K_3=K_1\cup\{\{b, c\}\}$, $K_4=K_2\cup K_3$, and $K_5=\S Even/(P)$.
It remains to show that the cells $K_0$ up to $K_5$ indeed occur as
the Kekul\'e cells of certain graphs.

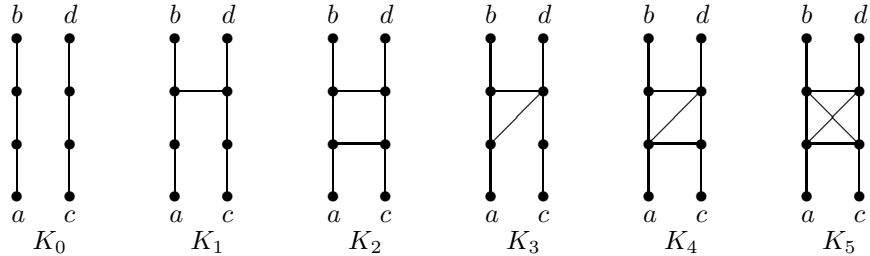
\begin{picture}(170,60)(0,-6)

\multiput(-5,0)(30,0){6}{
\put(20, 10){\line(0,1){30}}
\multiput(18.8,8.7)(0,10){4}{\bol}
\multiput(8.8,8.7)(0,10){4}{\bol}
\put(10, 10){\line(0,1){30}}
\put(9,5){$a$}
\put(19,5){$c$}
\put(9,43){$b$}
\put(19,43){$d$}}
\multiput(35,30)(30,0){5}{\line(1,0){10}}
\multiput(95,20)(30,0){3}{\line(1,1){10}}
\multiput(125,20)(30,0){2}{\line(1,0){10}}
\put(65,20){\line(1,0){10}}
\put(155,30){\line(1,-1){10}}
\put(8,0){$K_0$}
\put(38,0){$K_1$}
\put(68,0){$K_2$}
\put(98,0){$K_3$}
\put(128,0){$K_4$}
\put(158,0){$K_5$}
\end{picture}

\noindent 
It is easy to verify that the six graphs depicted above represent 
the Kekul\'e cells $K_0$ up to $K_5$, as required. \boks
\end{proof}

\begin{remark}
  This proof shows that the Kekul\'e cells with 4 ports and diameter 4
  can all be realized by subgraphs of the graph which is called
  $\Delta_4$ in section \ref{typeA}. It seems likely that this result
  can be generalized in some way, but at this point we do not see how.
\end{remark}

\section{Semi-Kekul\'e States} \label{semikekule}

This section serves as an aside to provide a context in which Kekul\'e
states are special cases. We weaken the condition for Kekul\'e states
to a linear condition over the field with two elements. This simplifies the
mathematics, since the whole theory of linear equations becomes
available. 

For an integer $x$, we write $x\B\ mod /2$ to denote the remainder of
$x$ upon integer division by 2. It follows that $x\B\ mod /2 = 0$ (or
$1$) if and only if $x$ is even (or odd, respectively). For integers
$x$ and $y$, we define $x\oplus y = (x+y)\B\ mod /2$.

A node in a graph is called \emph{even/odd} if its degree in the graph
is even/odd.  A subgraph $W$ of $G$ is called a \emph{semi-Kekul\'e}
state of $G$ if every internal node of $G$ is an odd node of
$W$. Clearly, every Kekul\'e state is a semi-Kekul\'e state, but not
\emph{vice versa}.

For any graph $G$, we define $\S odd/(G)$ to be the set of odd nodes
of $G$.  Therefore, $W$ is a semi-Kekul\'e state of $G$ if and only if
$iG\subseteq \S odd/(W)$. Since every edge contributes twice to the
degree of a node, the sum of the degrees of the nodes is twice the
number of edges. It is therefore even. Since the sum of the
degrees at the even nodes is also even, it follows that the sum of
the degrees of the odd nodes is even. Since every odd node contributes
an odd number, it follows that the number of odd nodes is even: $\#\S
odd/(G)\B\ mod /2 = 0$ for every graph $G$. 

\subsection{There are enough semi-Kekul\'e states}

Let the \emph{signature} of a graph $G$ be defined as the number $ sG
= (\#iG)\,\B mod /2$. In other words, the signature is 0 or 1,
depending on whether the number of internal nodes is even or odd.  To
formulate the next result conveniently, we define $\S Pow/(P)_0 = \S
Even/(P)$ and $\S Pow/(P)_1 = \S Odd/(P)$, see section \ref{abstr}.

\begin{theorem} \label{typeAmod2}
Let $G$ be a graph with port set $P=pG$ and signature $\eps=sG$.\\
(a) Let $W$ be a semi-Kekul\'e state of $G$. Then $(W|P)\in\S Pow/(P)_\eps$.\\
(b) Conversely, let $g\in \S Pow/(P)_\eps$.  Assume that graph $G$ is 
connected.  Then $G$ has a semi-Kekul\'e state $W$ with $(W|P) = g$.
\end{theorem}

\begin{proof}
For any subgraph $W$ of $G$, we define the set of violating nodes by
$\S vio/(W) = iG\setminus \S odd/(W)$. Clearly, $W$ is semi-Kekul\'e
if and only if $\S vio/(W)$ is empty.

On the other hand, we have $(W|P) = P\cap nW = \S odd/(W)\setminus iG$,
since ports are nodes that belong to precisely one edge. It follows that
\begin{tab}
\>\+ $ \#\S vio/(W) +\#(W|P) = \#(iG\oplus \S odd/(W)) $\\
$ =\; \# iG + \#\S odd/(W) - 2\cdot\#(iG\cap \S odd/(W)) $ .
\end{tab}
Since $\#\S odd/(W)$ is even and $\#iG\B\ mod /2 = \eps$, this implies that 
\begin{tab}
(*) \> $ (\#\S vio/(W)\B\ mod /2) \oplus (\#(W|P)\B\ mod /2) = \eps $ .
\end{tab}
If $W$ is semi-Kekul\'e, $\S vio/(W)$ is empty, so that $ \#(W|P)\B\
mod /2 = \eps $, i.e., $(W|P)\in \S Pow/(P)_\eps$. This proves part (a).

Conversely, assume that $\# g\B\ mod /2 = \eps$. We can define some
subgraph $W$ of $G$ with $(W|P)=g$.  Then the righthand summand of (*)
equals $\eps$. It follows that the number of elements of $\S vio/(W)$ is
even. If it is positive, then we can choose two different nodes $u$
and $v\in\S vio/(W)$. Since $V$ is connected, we can choose a simple
path $C$ from $u$ to $v$ in $G$, and replace $W$ by $W' = W\oplus C$,
which satisfies $\S vio/(W')=\S vio/(W)\setminus\{u,v\}$ and $W'|P =
g$. We can continue in this way until $\S vio/(W)$ is empty.  \boks
\end{proof}

\medbreak For a port assignment $g$ in a connected graph, Theorem
\ref{typeAmod2} implies that the condition $g\in \S Pow/(P)_\eps$ is
necessary and sufficient for the existence of a semi-Kekul\'e state
$W$ with $(W|P) = g$. This also implies that this is a necessary
condition for the existence of a Kekul\'e state $W$ with $(W|P) = g$.
In section \ref{typeA}, we will describe graphs in which this
condition is not only necessary but also sufficient.

\subsection{The homogeneous semi-Kekul\'e kernel}

Let \IB\ (pronounced \emph{Bit}) be the set of the two elements 0 and
1, with the addition $\oplus$ described above and an ordinary
multiplication. This is a field, i.e., an algebraic structure with the
usual laws for addition, subtraction, multiplication, and division by
nonzero elements. It follows that the whole theory of vector spaces
and linear equations is applicable over \IB .

In particular, for a set $P$, the additive group $\S Pow/(P)$ is a
vector space over \IB\ of dimension $\#P$. The easiest way to see this
is to identify a subset $g\subseteq P$ with its characteristic function
$\chi_g$ given by $\chi_g(p) = 1$ if $p\in g$ and $\chi_g(p) = 0$ if
$p\notin g$. In this way $\S Pow/(P)$ is identified with the set of
functions $P\to\IB$. The symmetric difference operator $\oplus$
corresponds to the operator on functions given by $(f\oplus g)(p) =
f(p)\oplus g(p)$.

The vector space $\S Pow/(G)$ is introduced in the same way.  We
define the \emph{homogeneous semi-Kekul\'e kernel}, $\S HSK/(G)$, of
$G$ to be the set of subgraphs $W$ for which all nodes are even:
\begin{tab}
\> $ \S HSK/(G) = \{W\in\S Pow/(G) \mid \S odd/(W)=\emptyset\}$ .
\end{tab}
It is easy to see that \S HSK/ is a linear subspace of $\S Pow/(G)$.

If an inhomogeneous system of linear equations has a solution $W_0$,
the set of all its solutions is of the form $W_0\oplus H$, where $H$
is the set of solutions of the homogeneous linear system. It follows
that, given one semi-Kekul\'e state $W_0$ of $G$, the set of
semi-Kekul\'e states with the same port assignment as $W_0$ is equal
to $W_0\oplus \S HSK/(G) = \{W_0\oplus W\mid W\in\S HSK/(G)\}$.

A graph is called a \emph{tree} if it is connected and has no
connected proper subgraph with the same set of nodes.  Every connected
graph has a spanning tree, i.e., a subgraph with the same nodes which
is a tree (just remove edges one by one until you would cause
disconnectedness).

\begin{theorem}
Let graph $G$ be connected. The vector space $\S HSK/(G)$ over $\IB$
has dimension $\#G+1-\#nG$. It has a basis over \IB\ that consists of
cycles of $G$.
\end{theorem}

\begin{proof}
  First, assume that $G$ is a tree. Then the number of nodes is one
  plus the number of edges, as is well-known in graph theory. So we
  have $\#nG=\#G+1$. On the other hand, since $G$ is a tree, one can
  prove by induction on the size of the tree that $\S
  HSK/(G)=\{\emptyset\}$.  This implies the assertion for the case
  that $G$ is a tree.

  We now proceed by induction on the number of edges. Let graph $G$ be
  given.  We may assume that $G$ is connected and not a tree.
  Therefore $G$ contains an edge $e$, such that removal of $e$ yields
  a connected graph $G'$. By induction, the space $\S HSK/(G')$ of the
  subgraphs $W$ of $G'$ with $\S odd/(W)$ empty has dimension $\#G
  -\#nG$ and a basis $B$ that consists of cycles of $G'$. Since graph
  $G'$ is connected, edge $e$ is contained in a cycle $C$ of graph
  $G$.  Now $C$ is an element of $\S HSK/(G)\setminus \S HSK/(G')$.
  Therefore $\S HSK/(G')$ is a proper subset of $\S HSK/(G)$. Since
  $\S HSK/(G')$ is the subset of $\S HSK/(G)$ given by the single
  equation that corresponds to the condition $e\notin W$, it follows
  that $\S dim/(\S HSK/(G)) = 1+\S dim/(\S HSK/(G'))=\#G+1-\#nG$ and
  that $B\cup\{C\}$ is a basis of $\S HSK/(G)$. \boks
\end{proof}

\medbreak This result was inspired by the observation that, for each
of the graphs considered in Figures 5.7 and 5.8 of \cite{Vee06}, the
set $\S HSK/(G)$ has 8 elements.

\begin{corollary} \label{basHSK}
Let $G$ be connected. Let $C_0$, \dots , $C_{r-1}$ be cycles in
$G$ such that $r=\#G+1-\#nG$ and that 
$C_0$, \dots , $C_{r-1}$ are linearly independent over \IB .\\
(a) Then $C_0$, \dots , $C_{r-1}$ form a basis of $\S HSK/(G)$.\\
(b) Let $P=pG$ and $\eps = sG$. For any $g\in\S Pow/(P)_\eps$, there 
are precisely $2^r$ semi-Kekul\'e states $W$ with $(W|P)=g$.
\end{corollary}

From part (b) it follows that the number of Kekul\'e states with a 
given port assignment is $\leq 2^r$.

Usually, it is very easy to find enough linearly independent
cycles to obtain a basis for \S HSK/ with Corollary \ref{basHSK}, and to
find simple paths between different ports. Since the constructive
proof of Theorem \ref{typeAmod2} is not very efficient, we prefer to
apply Gauss elimination to the inhomogeneous system to find
semi-Kekul\'e states.

\B Complexity./ The problem to determine some Kekul\'e state for a
given graph with given port assignment, has a complicated but
efficient solution with complexity $\Oh(\#E\cdot\sqrt{\#V})$
\cite{MiV80}. All semi-Kekul\'e states can easily be found in time
$\Oh(\#E\cdot\#V)$. The above approach leads to an algorithm to
determine all Kekul\'e states in time $\Oh(\#E\cdot(\#V+2^r))$ where
$r$ is as in Corollary \ref{basHSK}. In the applications, this is
quite feasible, since $r$ tends to be rather small.

\section{Omniconjugated Graphs} \label{typeA}

Van der Veen et al.\ \cite{VRJH04} have defined $n$-port
(`$n$-terminal') $\pi$-conjugated molecules, with $n\geq 3$, as
omniconjugated of type A, if direct linear conjugated pathways are
present between all pairs of ports in all states of the ports. We now
formalize the principles of such systems in a mathematical way.

We define a graph $G$ with signature $\eps$ to be \emph{omniconjugated}
if $P=pG$ has at least two elements and the Kekul\'e
cell of $G$ is $\S Pow/(P)_\eps$ (i.e., maximal). Note that our concept 
omniconjugated represents what is called ``omniconjugated of Type A'' 
in \cite{VRJH04,Vee06}.

\begin{corollary} \label{thmApath}
  Let $G$ be omniconjugated. Let $W$ be a Kekul\'e state of $G$.
  Let $p$ and $q$ be two different ports of $G$.  Then $G$ has an
  alternating curve $(C,W)$, such that $C$ is a simple path from $p$
  to $q$.
\end{corollary}

\begin{proof}
  Since $(W|P)\oplus \{p,q\} \in\S Pow/(P)_\eps$, and the graph is
  omniconjugated, the assertion follows from Theorem \ref
  {portchannels}.  \boks
\end{proof}

\medbreak The corollary means that all channels of an omniconjugated
graph are always open (low resistance). Conversely, if graph $G$ has
at least one Kekul\'e state and all channels are always open, it is 
omniconjugated.

For every set $P$ with at least two elements, omniconjugated graphs
exist. This is shown as follows. Let $A_2$ be the graph that consists
of one edge with its two nodes. The two nodes are ports. In this case,
$\eps = 0$ and $\S Pow/(P)_\eps$ contains the empty subgraph and the
full subgraph which are both Kekul\'e states. Therefore $A_2$ is 
omniconjugated. More generally, the linear graph $A_n$ of $n\geq 2$
consecutive nodes and $n-1$ edges is also omniconjugated (the name $A_n$
comes from the classification of Dynkin diagrams).

The smallest omniconjugated graph with $\geq 3$ ports is the graph
$\Delta_3$ with 3 ports, and three internal nodes of degree 3 that
form a triangle.  More generally, let $G_n$ for $n\geq 2$ be the
complete graph with $n$ nodes and ${n \choose 2} $ edges. Let
$\Delta_n$ be the graph obtained from $G_n$ by attaching a port via an
edge to every node of $G_n$. So, $\Delta_n$ has $2n$ nodes and ${n
\choose 2}+n$ edges.  Note that $\Delta_2=A_4$ and that $\Delta_4$ is
the graph used for $K_5$ in the proof of Lemma \ref{class4}. We
leave it to the reader to verify that all graphs $\Delta_n$ with
$n\geq 2$ are omniconjugated. Conversely, we have:

\begin{lemma} \label{delta}
  Let $G$ be a graph with $n\geq 2$ nodes. Let graph $G'$ be
  obtained from $G$ by attaching a port via an edge to every node of
  $G$. Assume that $G'$ is omniconjugated. Then $G$ is a complete graph.
\end{lemma}

\begin{proof}
Since $G$ has no isolated nodes, all nodes of $G$ are internal nodes of
$G'$, and $G'$ has $n$ ports. Therefore, the signature $\eps$ of
$G'$ is $n\B\ mod /2$. Let $P$ be the set of ports of $G'$.

It suffices to prove that every pair of different nodes of $G$ is
linked in $G$. Let $u\ne v$ be nodes of $G$. Let $p$ and $q$ be the
unique ports of $G'$ linked to $u$ and $v$, respectively.  Let $g$ be
the port assignment of $G'$ given by $g=P\setminus \{p,q\}$.  We have
$g\in \S Pow/(P)_\eps$ since $(n-2)\B\ mod /2=\eps$.

Since it is omniconjugated, graph $G'$ has a Kekul\'e state $W$ with
$(W|P)=g$.  Every edge $e$ that contains a port $r\in g$ is in $W$.
Since $W$ is a Kekul\'e state of $G'$, it follows that $e\notin W$ for
every edge $e\in G$ that contains some node of $G$ other than $u$ and
$v$. The two edges $e$ of $G'$ incident with $p$ and $q$ are not in
$W$.  Since $W$ is a Kekul\'e state of $G'$, it follows that $u$ is
element of a unique edge $e$ of $W$.  It follows that $v$ is the other
endpoint of $e$. This proves that $u$ and $v$ are linked in $G$. \boks
\end{proof}

\medbreak Section 3.3.1 of \cite{Vee06} presents a topological design
program for omniconjugated models that have properties similar to
those of our omniconjugated graphs. We transfer this idea to
omniconjugated graphs in the following way.

Starting with an omniconjugated graph $G$, the question is: what kinds
of operations on $G$ are allowed that preserve its property of being
omnicojugated? It turns out that operations like \B Op.vi/ and \B
Op.vii/ of \cite{Vee06} are allowed because of Theorem
\ref{mergenodes}. It is also easy to see that the operations \B
Op.ii/, \B Op.v/, and \B Op.iv/ are preserved in the following sense:

\begin{lemma} \label{opii+v}
  Let $G$ be an omniconjugated graph. Let $G'$ be obtained from $G$ by
  applying
  the following two types of operations:\\
  \B Op.ii./ Add arbitrary edges between internal nodes.\\
  \B Op.v./ Replace the edge to a port by two edges and a node of degree 2.\\
  Then $G'$ is omniconjugated.
\end{lemma}

\begin{lemma} \label{opiv}
\B Op.iv./ Let $G'$ be a graph with port $p'$ and let $G''$ be a graph
with port $p''$. Let graph $G$ be obtained from $G'$ and $G''$ by
removing the two ports and identifying their incident edges. Then $G$
is omniconjugated if and only if $G'$ and $G''$ are omniconjugated.
\end{lemma}

For example, as indicated in \cite[Scheme 3.1]{Vee06}, we can apply
\B Op.ii/ twice to the basic model $A_6$ to obtain the basic model \B
B/, which is therefore also omniconjugated.

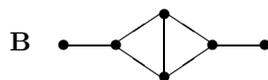
\begin{picture}(120,20)(0,0)
\put(12,9){\B B/}
\put(41.4,4.2) {\line(0,1){11.8}}
\multiput(40.2,2.8)(0,12){2} {\bol}
\put(49.8,10) {\line(1,0){10}}
\multiput(49.3,8.8)(10,0){2} {\bol}
\put(21.5,10) {\line(1,0){10}}
\multiput(21,8.8)(10,0){2} {\bol}
\put(42,4.2) {\line(3,2){8}}
\put(42,16) {\line(3,-2){8}}
\put(40.6,4.2) {\line(-3,2){8}}
\put(40.6,16) {\line(-3,-2){8}}
\end{picture}

Starting with graph $\Delta_3$, one can apply \B Op.iv/ to obtain
omniconjugated graphs with arbitrary many ports. For this purpose, we
therefore do not need the graphs $\Delta_n$ with $n>3$.  

In Theorem \ref{mergenodes}, we described a transformation from a
graph $G$ into a graph $G'$ that has the same Kekul\'e cell. It
follows that $G'$ is omniconjugated if and only if $G$ is
omniconjugated.  We conjecture that every omniconjugated graph in
which all nodes have degree $\leq 3$ can be obtained by starting with
the basic models $A_2$ and $\Delta_3$ and applying Theorem
\ref{mergenodes} and the Lemmas \ref {opii+v} and \ref{opiv}.

\section{Functional Cells and Controlled Switches} \label{switch}

In this section, we build upon the ideas of Chapter 6 of \cite{Vee06}.
The idea of this chapter is that an alternating path between a pair of
ports forms an open channel, and that the absence of any alternating
path means closure of the channel.

A cell $K\subseteq \S Pow/(P)$ becomes a \emph{functional cell} by
specifying an initial element $k_0\in K$ and a linearly independent
system of channels to probe and modify the current element of the
cell. See the example given at the end of section \ref{portass}.

\subsection{A cell for the conjunction}

Section 6.3 of \cite{Vee06} contains 15 of the 16 binary boolean
operators implemented with pyracylene-based structures bearing six
ports in various patterns.  The only missing operator is the
conjunction. Since $\neg A\land B$ is available in structure \B 6.8/
(p.\ 140) of \cite{Vee06}, the conjunction can be realized in
pyracylene by taking a nonzero initial state.

We now sketch an approach that yields a simpler graph.  We first
design the functional cell. Let $k_0=\emptyset$ be the initial
element. In the style of \cite{Vee06}, we choose two input channels
$A$ and $B$ and a test channel $T$. Since we want to model the
conjunction of the inputs, test channel $T$ must be open if and only
if both input channels are signalled. Therefore, the cell $K$ must
contain the elements $\emptyset$, $A$, $B$, $A\oplus B$, and $A\oplus
B\oplus T$.  Since channel $T$ must be closed in the states $\emptyset$,
$A$, $B$, cell $K$ must not contain the elements $T$, $A\oplus T$, and
$B\oplus T$.

Since all eight port assignments mentioned should be different, the 
smallest candidate is a cell with 4 ports and 5 elements. The solution
happens to occur as cell $K_1$ in the classification in the proof of
Lemma \ref{class4} when we take $A=\{a, b\}$, $B=\{c, d\}$, and
$T=\{b, d\}$.

A disadvantage of this solution is that both ports of test channel $T=
\{b, d\}$ are used in the input channels. It may be difficult to
independently signal $A$ and $B$ and probe $T$.

\subsection{Reading output without disabling the switch}

The example discussed above has an even more fundamental problem,
which it shares with the example in section \ref{portass} and the
results of \cite{Vee06}: if the test whether the output channel $T$ is
open is performed by signalling the channel, the cell moves to a state
where it no longer satisfies its specification.

The point is discussed by \cite{Vee06} on page 149. She distinguishes
signalling a channel by sending a soliton through it and reading a
channel by means of a polaron. A polaron can be thought of as a double
soliton \cite{RoB87}. In order to use the output of one cell as an
input of another cell, we would like to read the output by means of
solitons.  This, however, requires that signalling the output channel
leaves the cell in a state equivalent to the state before the output
was read.

In the start state, and in all states reachable before the output is
read, all inputs should be possible. In some but not all of these
states, the output channel $T$ should be open. This implies that
signalling $T$ when it is open, can bring the cell in a state where
some of the input channels are not open. We therefore have to relax
the requirement that input channels are always open, but since input
must always be possible, we have to share inputs over more than one
channel.

We therefore define an input \emph{socket} to consist of a pair
$(A,B)$ of parallel channels of the form $A=\{p, r\}$, $B=\{q, r\}$
for ports $p$, $q$, and $r$, such that always either $A$ is open and
$B$ is closed, or \emph{vice versa}.  When a soliton is sent to a socket, it
will travel through an open channel of the two.

\begin{remark} 
Note that we require that the two parallel channels of the socket
share a port. In fact, if the parallel channels $A$ and $B$ are
disjoint, a soliton sent through the pair can enter via a port of $A$
and exit via a port of $B$, so that an unintended channel is
signalled. In order to reckon with this possibility, one would have to
impose additional requirements. \boks
\end{remark}

\begin{example} The simplest case, called the \emph{Y-cell}, has one
  input socket and one output channel. Let the socket consist of the
  channels $A$ and $B$, and let $T$ be the output channel. The vectors
  $A$, $B$, $T$ are supposed to be linearly independent. We postulate
  that $K=k_0\oplus L$ where $L=\{\emptyset, A, A\oplus T, A\oplus
  B\oplus T\}$.
 
The state space $K$ is a translation of $L$. Either space consists of
4 states with three possible transitions. In the diagram, a transition
indicated with channel $c$ corresponds to the operation $x\mapsto x\oplus c$.

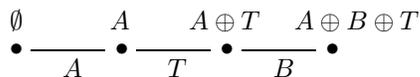
\begin{picture}(120,20)(6,3)
\put(20,15) {$\emptyset$}
\put(39,15) {$A$}
\put(54,15) {$A\oplus T$}
\put(74,15) {$A\oplus B\oplus T$}
\multiput(20,10)(20,0){4} {\bol}
\multiput(24,11)(20,0){3} {\line(1,0){14}}
\put(30,6) {$A$}
\put(50,6) {$T$}
\put(70,6) {$B$}
\end{picture}

\noindent In all states, either channel $A$ or channel $B$ is
available, and toggling this channel opens or closes the output
channel $T$. 

The simplest graph with such a Kekul\'e cell is a tree with 5 ports
and 3 internal nodes. In this case, the vector space $\S Pow/(P)_\eps$ 
over \IB\ has dimension 4 (16 elements). The Kekul\'e cell has 8
elements. We use the initial state and the input socket $(A,B)$ and
the output channel $T$ as depicted in the leftmost graph.

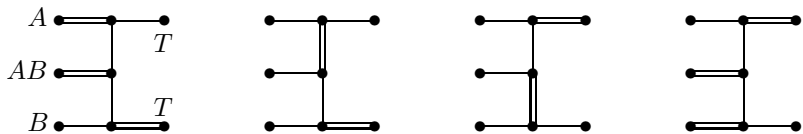
\begin{picture}(160,35)(15,-2)
\multiput(0,0)(40,0){4}{
\multiput(28.7,3.7)(0,10){3} {\bol}
\multiput(38.7,3.7)(0,10){3} {\bol}
\multiput(48.7,3.7)(0,20){2} {\bol} }
\multiput(30,24.5)(0,1){2}{\line(1,0){10}}
\multiput(30,14.5)(0,1){2}{\line(1,0){10}}
\multiput(0,0)(40,0){2}{\multiput(40,4.5)(0,1){2}{\line(1,0){10}}}
\multiput(30,5)(40,0){3}{\line(1,0){10}}
\multiput(40,25)(40,0){2}{\line(1,0){10}}
\multiput(40,5)(120,0){2}{\line(0,1){20}}
\put(24,24) {$A$}
\multiput(48,7)(0,12){2} {$T$}
\put(20,14) {$AB$}
\put(24,4) {$B$}
\multiput(70,25)(40,0){3}{\line(1,0){10}}
\multiput(70,15)(40,0){2}{\line(1,0){10}}
\multiput(120,5)(40,0){2}{\line(1,0){10}}
\multiput(80,5)(40,10){2}{\line(0,1){10}}
\multiput(0,0)(40,-10){2}{\multiput(79.5,15)(1,0){2}{\line(0,1){10}}}
\multiput(0,0)(40,0){2}{\multiput(120,24.5)(0,1){2}{\line(1,0){10}}}
\multiput(0,0)(0,10){2}{\multiput(150,4.5)(0,1){2}{\line(1,0){10}}}
\end{picture}

\noindent In the initial state depicted on the left, only channel $A$
is open. If channel $A$ is signalled, we get in state middle-left
where the channels $A$ and $T$ are open. If $T$ is signalled there, we
get in state middle-right where the channels $B$ and $T$ are open. If
$B$ is signalled there, we get in the rightmost state where channel
$B$ is the only open channel.

This state of affairs can also be realized in a pyracylene derivative
as shown below. The ports $A_1$ and $T_1$ start with a double bond in
the initial state $k_0$.

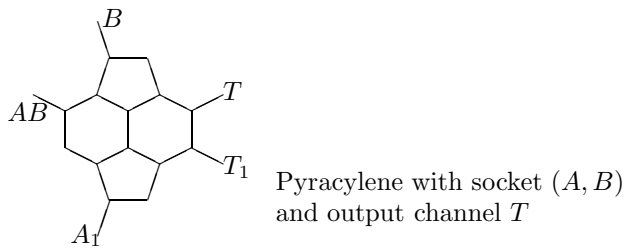
\begin{picture}(120,50)(-10,12)
\multiput(20,40)(12,0){3} {\line(2,1){6}}
\multiput(14,43)(12,0){3} {\line(2,-1){6}}
\multiput(20,40)(12,0){3} {\line(0,-1){7}}
\multiput(20,33)(12,0){3} {\line(2,-1){6}}
\multiput(26,30)(12,0){2} {\line(2,+1){6}}
\put(21,15) {$A_1$}
\put(9,38) {$AB$}
\put(27,56) {$B$}
\put(50,42) {$T$}
\put(50,28) {$T_1$}
\put(60,25) {Pyracylene with socket $(A,B)$}
\put(60,19) {and output channel $T$}
\put(26,43)   {\line(1,3){2.3}}
\put(38,43)   {\line(-1,3){2.3}}
\put(28.5,50) {\line(1,0){7}}
\put(28.5,50) {\line(-1,3){2.3}}
\put(26,30)   {\line(1,-3){2.3}}
\put(38,30)   {\line(-1,-3){2.3}}
\put(28.5,23) {\line(1,0){7}}
\put(28.5,23) {\line(-1,-3){2.3}}
\end{picture}

\noindent
In this case, $\S Pow/(P)_\eps$ has
dimension 4. The Kekul\'e cell has 12 elements. From the start state
$k_0$ only four of them are reachable by sending solitons over the
channels $A$, $B$, $T$, but, e.g., one reaches the origin $\emptyset$ by
sending a soliton from $A_1$ to $T_1$.  The origin has four different
Kekul\'e states, whereas $k_0$ and $k_0\oplus A\oplus B\oplus T$ have
only two different Kekul\'e states.  This suggests that the origin is
physically much more stable than the physical states corresponding to
$k_0$ and $k_0\oplus A\oplus B\oplus T$.  The port assignments
$k_0\oplus A$ and $k_0\oplus A\oplus T$ have only one Kekul\'e
state. It is likely that such Kekul\'e states represent less stable physical states. \boks
\end{example}

\begin{remark}
  Is there a graph $G$ with four ports $p$, $q$, $r$, $t$, such that
  the above structure can be realized by taking $A=\{p, r\}$,
  $B=\{q, r\}$, $T=\{t, r\}$? The answer is no. This follows from
  Theorem \ref{kekConByChan} by taking $g=\emptyset$ and $ g'= A\oplus
  B\oplus T$. \boks
\end{remark}

\subsection{Splitting for feedback}

One may decide that, when an output channel $T$ is read and toggled, a
signal is sent back through $T$ to reset the cell to its unread
state. Since the output should also be sent forward to be used, we
need a splitter: a cell with one input channel or socket and two
output channels that are both opened when an input signal arrives.

We use an input socket $(A,B)$ and output channels $S$ and $T$. Let
the origin $\emptyset$ be the initial state. Then the cell must contain \B
0/, $A$, $A\oplus S$, $A\oplus S\oplus T$, $A\oplus S\oplus T\oplus
B$.  Since the channels $S$ and $T$ must be closed in states $\emptyset$ and
$A\oplus S\oplus T\oplus B$, the cell must not contain $S$, $T$,
$A\oplus S\oplus B$, and $A\oplus T\oplus B$.  Since $(A,B)$ is a
socket, the cell must not contain $B$, $A\oplus B$, $S\oplus T$, and
$S\oplus T\oplus B$. The cell may or may not contain the elements
$A\oplus T$, $B\oplus S$, and $B\oplus T$.

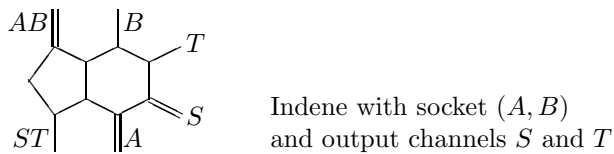
\begin{picture}(120,36)(-10,20)
\multiput(20,40)(12,0){2} {\line(2,1){6}}
\multiput(14,43)(12,0){2} {\line(2,-1){6}}
\multiput(20,40)(12,0){2} {\line(0,-1){7}}
\put(20,33) {\line(2,-1){6}}
\multiput(14,30)(12,0){2} {\line(2,+1){6}}
\multiput(31.8,32.5)(.5,.7){2} {\line(2,-1){6}}
\put(27,24) {$A$}
\put(5,46) {$AB$}
\put(6,24) {$ST$}
\put(27,46) {$B$}
\put(39,42) {$T$}
\put(39,28) {$S$}
\put(55,30) {Indene with socket $(A,B)$}
\put(55,24) {and output channels $S$ and $T$}
\put(14,43)   {\line(-3,-4){5}}
\put(14,30)   {\line(-3,4){5}}
\multiput(13.6,43)(1,0){2}{\line(0,1){7}}
\multiput(25.5,30)(1,0){2}{\line(0,-1){7}}
\put(26,43){\line(0,1){7}}
\put(14,30){\line(0,-1){7}}
\end{picture}

\noindent 
The indene molecule has a graph with a Kekul\'e cell with these
properties, as depicted here. The 9 interior nodes form a pentagon and
a hexagon. In this case $\S Pow/(P)_\eps$ has dimension 5, and 32 elements.
The Kekul\'e cell has 18 elements. We use the input socket $(A,B)$ and
the output channels $S$, $T$ as depicted. Incidentally, $S$ and $T$
also form a socket.  Initially, there are double bonds at both ports
of channel $A$ and at the eastern port of $S$. Indene has
several solutions of this type, but all of these have initially three
double bonds at ports, have output channels combined in a socket, and
have the elements $A\oplus T$, $B\oplus S$, and $B\oplus T$ in the
Kekul\'e cell.

\section{Conclusions} \label{conclusions}

In this paper, we presented a theoretical investigation into the
computational possibilities of polycyclic hydrocarbons, based on
connectivity via molecular paths with alternating single and double
bonds. We confirmed and extended the results presented in
\cite{Vee06}. Moreover, we introduced the new concepts of cell and
functional cell in order to characterize the switching behaviour of
single unsaturated polycyclic hydrocarbon molecules. We regard this as
an interesting and fascinating subject, both from the point of view of
new chemistry and new applications of mathematics, as well as for the
speculations about their potential for molecular computation.  We
showed that it is, in principle, possible to build logical switches
with $\pi$-conjugated systems of unsaturated polycyclic hydrocarbons,
but we are fully aware of the tremendous technical problems in
realizing this potential. Anyway, the notion of functional cell
emerging from our work is, in its own right, an interesting starting
point for an alternative abstract model for computation. We showed
that it encompasses logical switches; it would be interesting to
investigate whether and how, e.g., data storage can be modeled with
it.

So far, we have not yet found a connection between
our investigations and quantum computing, although such a connection
might very wel exist. The relevant properties of unsaturated
polycyclic hydrocarbons are a direct consequence of the quantum
mechanical behaviour of electrons, but there seems no role for
entanglement here, which renders the connection with quantum computing
still elusive.


\begin{thebibliography}{10}

\bibitem{Ca82}
F.L. Carter.
\newblock {\em Molecular Electronic Devices}.
\newblock Marcel Dekker, 1982.

\bibitem{Cou61}
C.A. Coulson.
\newblock {\em Valence}.
\newblock Oxford Press, London, 1961.

\bibitem{DMKao03}
F.~Duli\'c, S.J. van~der Molen, T.~Kudernac, H.T. Jonkman, J.J. de~Jong, T.N.
  Bowden, J.~van Esch, B.L. Feringa, and B.J. van Wees.
\newblock One-way optoelectronic switching of photochromic molecules on gold.
\newblock {\em Phys. Rev. Lett.}, 91:207402, 2003.

\bibitem{FoR02}
P.W. Fowler and A.~Rassat.
\newblock Symmetry and distortive $\pi$-electrons in two- and three-dimensional
  conjugated systems.
\newblock {\em Phys. Chem. Chem. Phys.}, 4:1105--1113, 2002.

\bibitem{GLS88}
M.\ Gr\"otschel, L.\ Lov\'asz, and A.\ Schrijver.
\newblock {\em Geometric algorithms and combinatorial optimization}.
\newblock Springer V., Berlin, etc., 1988.

\bibitem{HKSS88}
A.J. Heeger, S.~Kivelson, J.R. Schrieffer, and W.-P Su.
\newblock Solitons in conducting polymers.
\newblock {\em Rev. Mod. Phys.}, 60:781, 1988.

\bibitem{HeL27}
H.~Heitler and F.~London.
\newblock Wechselwirkung neutraler atome und hom\"opolare bindung nach der
  quantenmechanik.
\newblock {\em Z. Phys.}, 44:455, 1927.

\bibitem{Lew16}
G.N. Lewis.
\newblock The atom and the molecule.
\newblock {\em J. Am . Chem. Soc.}, 38:762, 1916.

\bibitem{MWREao03}
M.~Mayor, H.B. Weber, J.~Reichert, M.~Elbing, C.~Von H\"anish, D.~Beckmann, and
  M.~Fisher.
\newblock Electric current through a molecular rod-relevance of the position of
  the anchor groups.
\newblock {\em Angew. Chem. Int. Ed.}, 42:5834, 2003.

\bibitem{MiV80}
S.~Micali and V.V. Vazirani.
\newblock An {${\cal O}(\sqrt{|v|}\cdot |E|)$} algorithm for finding maximum
  matching in general graphs.
\newblock In {\em 21st Annual Symposium on Foundations of Computer Science},
  pages 17--27. IEEE, 1980.

\bibitem{NiR03}
A.~Nitzan and M.A. Ratner.
\newblock Electron transport in molecular wire junctions.
\newblock {\em Science}, 300:1384, 2003.

\bibitem{OSR01}
S.~Owre, N.~Shankar, J.M. Rushby, and D.W.J. Stringer-Calvert.
\newblock {\em PVS Version 2.4, System Guide, Prover Guide, PVS Language
  Reference}, 2001.
\newblock \verb!http://pvs.csl.sri.com!

\bibitem{Pog00}
L.~Pogliani.
\newblock From molecular connectivity indices to semiempirical connectivity
  terms: recent trends in graph theoretical descriptors.
\newblock {\em Chem. Rev.}, 100:3827--3858, 2000.

\bibitem{RoB87}
S.~Roth and H.~Bleier.
\newblock Solitons in polyacetylene.
\newblock {\em Adv. Phys.}, 36:385, 1987.

\bibitem{ShH03}
S.~Shaik and P.C. Hiberty.
\newblock Myth and reality in the attitude toward valence-bond ({VB}) theory:
  Are its failures real?
\newblock {\em Helv.\ Chim.\ Acta}, 86:1063--1084, 2003.

\bibitem{SSZ95}
I.V. Stankevich, M.I. Skvortsova, and N.S. Zefirov.
\newblock On a quantum chemical interpretation of molecular connectivity
  indices for conjugated hydrocarbons.
\newblock {\em J. Molecular Structure}, 342:173--179, 1995.

\bibitem{Vee06}
M.H.\ van~der Veen.
\newblock {\em $\pi$-Logic}.
\newblock PhD thesis, University of Groningen, May 2006.

\bibitem{VRJH04}
M.H. van~der Veen, M.T. Rispens, H.T. Jonkman, and J.C. Hummelen.
\newblock Molecules with linear $\pi$-conjugated pathways between all
  substituents: Omniconjugation.
\newblock {\em Adv. Funct. Mat.}, 14:215--223 and 742, 2004.

\bibitem{vDMvdVH06}
E.H. van Dijk, D.J.T. Myles, M.H. van~der Veen, and J.C. Hummelen.
\newblock Synthesis and properties of an anthraquinone-based redox switch for
  molecular electronics.
\newblock {\em Organic Letters}, 8:2333--2336, 2006.

\bibitem{YaR02}
S.N. Yalirahi and M.A. Ratner.
\newblock Interplay of topology and chemical stability on the electronic
  transport of molecular junctions.
\newblock {\em Ann. N.Y. Acad. Sci.}, 960:153, 2002.

\end{thebibliography}
\end{document}